\documentclass[prx,twocolumn,floatfix,superscriptaddress, longbibliography]{revtex4-1}
\usepackage[pdftex,plainpages=false,colorlinks=true,linkcolor=blue, citecolor=blue, urlcolor=blue]{hyperref}
\usepackage{amsfonts}
\usepackage{amsmath}
\usepackage{amssymb}
\usepackage{graphicx}% Include figure files
\usepackage{natbib}
\usepackage{color}
\usepackage{placeins}
\usepackage{physics}

\begin{document}
 
\title{Critical Opalescence across the Doping Driven Mott Transition in Optical Lattices of Ultracold Atoms}
\author{C. Walsh}
\affiliation{Department of Physics, Royal Holloway, University of London, Egham, Surrey, UK, TW20 0EX}
\author{P. S\'emon}
\affiliation{Computational Science Initiative, Brookhaven National Laboratory, Upton, NY 11973-5000, USA}
\author{G. Sordi}
\email[corresponding author: ]{giovanni.sordi@rhul.ac.uk}
\affiliation{Department of Physics, Royal Holloway, University of London, Egham, Surrey, UK, TW20 0EX}
\author{A.-M. S. Tremblay}
\affiliation{D\'epartement de physique \& Institut quantique \& RQMP, Universit\'e de Sherbrooke, Sherbrooke, Qu\'ebec, Canada J1K 2R1}
\affiliation{Canadian Institute for Advanced Research, Toronto, Ontario, Canada, M5G 1Z8}

\date{\today}

\begin{abstract}
Phase transitions and their associated crossovers are imprinted in the behavior of fluctuations. Motivated by recent experiments on ultracold atoms in optical lattices, we compute the thermodynamic density fluctuations $\delta N^2$ of the two-dimensional fermionic Hubbard model with plaquette cellular dynamical mean-field theory. To understand the length scale of these fluctuations, we separate the local from the nonlocal contributions to $\delta N^2$. We determine the effects of particle statistics, interaction strength $U$, temperature $T$ and density $n$. At high temperature, our theoretical framework reproduces the experimental observations in the doping-driven crossover regime between metal and Mott insulator. At low temperature, there is an increase of thermodynamic density fluctuations, analog to critical opalescence, accompanied by a surprising reduction of the absolute value of their nonlocal contributions. This is a precursory sign of an underlying phase transition between a pseudogap phase and a metallic phase in doped Mott insulators, which should play an important role in the cuprate high-temperature superconductors. Predictions for ultracold atom experiments are made. 
\end{abstract}
 
\maketitle

\section{Introduction}
\label{S:Introduction}

Theoretical models that can give, with minimal assumptions, insights on how complex behaviors and structures emerge in natural phenomena are particularly appealing. The Hubbard model for strongly interacting electrons on a lattice is such a model. It contains the competition between a kinetic energy term, that describes the delocalisation of particles on a lattice, and an on-site repulsive interaction $U$ term, that favors instead particle localisation. As a prominent example, this model has long helped to describe the Mott transition~\cite{mott, ift}, where electron-electron repulsion turns a metal into an insulator. It has also helped to throw light on how complex phases of matter in quantum materials, such as cuprate high-temperature superconductors~\cite{Anderson:1987, LeeRMP:2006, tremblayR}, arise from strong interactions. 

Nevertheless, correlated electron systems pose fundamental challenges to many-particle quantum theory since even a minimal model such as the Hubbard model is hard to solve or simulate~\cite{Troyer:2005}. Ultracold atoms in optical lattices have emerged as a useful platform for implementing the Hubbard model and thereby for performing ``quantum simulations'' of interacting quantum systems~\cite{GrossScience2017, bdzRMP, jzHM, lewenstein, Esslinger:2010}. The synergy between advances such as single atom microscopy~\cite{Bakr:2009, Sherson:2010, Cheuk:PRL2015, Haller:NatPhys2015, Parsons:PRL2015, Omran:PRL2015, Edge:PRA2015} in ultracold atom experiments and theoretical research is leading to a better understanding of the phase diagram of the Hubbard model. Important milestones include the observation of the Mott insulating phase~\cite{Jordens:2008, Schneider:2008, Cheuk:PRL2016, Greif:Science2016}, the measure of the equation of state~\cite{Cocchi:PRL2016, Hofrichter:PRX2016}, the detection of short-range spin and charge correlations~\cite{Cheuk:2016, Parsons:2016, Boll:2016, DrewesPRL2016, drewesPRL2017} and their implication for transport properties~\cite{Nichols:2018, Brown:Science2019}, the measure of entanglement~\cite{greinerNat2015, Cocchi:PRX2017, walshSl, walshSb}, and the observation of long-range antiferromagnetism~\cite{mazurenko2017cold}.

A central programme of the synergic effort in theory and experiments with ultracold atoms is the investigation of the Hubbard model in two dimensions. This is motivated by the link to the physics of cuprate high-temperature supercondutors. These materials are doped Mott insulator whose important properties should be captured by this model. 

In cold atoms investigations, density fluctuations measured through single-atom microscopy have emerged as a powerful tool~\cite{Muller:2010, Sanner:2010, DrewesPRL2016} to detect and characterize phase transitions and crossovers, and hence to throw new light on the phase diagram of the Hubbard model. For example, experiments in Ref.~\onlinecite{DrewesPRL2016} compared thermodynamic and local density fluctuations, unveiling a clear metal to Mott insulator crossover as a function of particle density (or doping): in the Mott insulating state thermodynamic and local fluctuations are both suppressed, whereas in the metallic state density fluctuations of all length scale, and not only local ones, contribute to the thermodynamic fluctuations. 

What happens to this metal to Mott insulator crossover at much lower temperatures? While such temperatures are currently inaccessible experimentally, this may change soon. Similarly, what happens to crossovers as a function of doping? The answer to this question bears key implications for the physics of the cuprates. Upon doping the Mott insulator, a pseudogap phase emerges. The nature of this phase is under intense scrutiny in cuprates. Based on recent theoretical work~\cite{sht, sht2, ssht} with the cellular extension~\cite{maier, kotliarRMP, tremblayR} of dynamical mean-field theory~\cite{rmp} (CDMFT), the key qualitative feature of the temperature-doping phase diagram of the two-dimensional Hubbard model is a finite-temperature and finite-doping first-order transition between a metal at low density and a pseudogap at high density~\cite{sht, sht2}, which ends in a second-order critical endpoint from which a sharp crossover, called Widom line, emerges up to high temperatures~\cite{ssht}. The defining feature of the critical endpoint is the divergence of the correlation length $\xi$, while the Widom line is the locus of the maxima of $\xi$ in the supercritical region. Now, density fluctuations are sensitive to $\xi$, and hence they are expected to be sensitive signatures of both the endpoint and of the associated Widom line. 

These signatures are the focus of this paper. We calculate separately the local and nonlocal contributions to density fluctuations for the square lattice near-neighbor two-dimensional Hubbard model using CDMFT, spanning two decades in temperatures, ranging from temperatures accessible to experiment down to the lower temperatures needed to detect the first-order transition. In Section~\ref{S:FluctHighT} we first analyze the density fluctuations at temperatures accessible to experiments, and then compare them with the available experimental data of Ref.~\onlinecite{DrewesPRL2016}. The excellent agreement between theory and experiment proves the accuracy of our approach and provides a strong framework to interpret current experiments. In Section~\ref{S:FluctLowT}, we then compute the density fluctuations at low temperatures, making predictions in Sec.~\ref{S:Prediction} that may allow experiments to detect the Widom line, critical endpoint and first-order transition. The low-temperature predictions are robust but qualitative since a highly frustrated lattice should be used to avoid magnetic phase transitions. The fermion sign problem prevents us from computing on such lattices.

\begin{figure*}%[h!]
\centering{
\includegraphics[width=1.\linewidth]{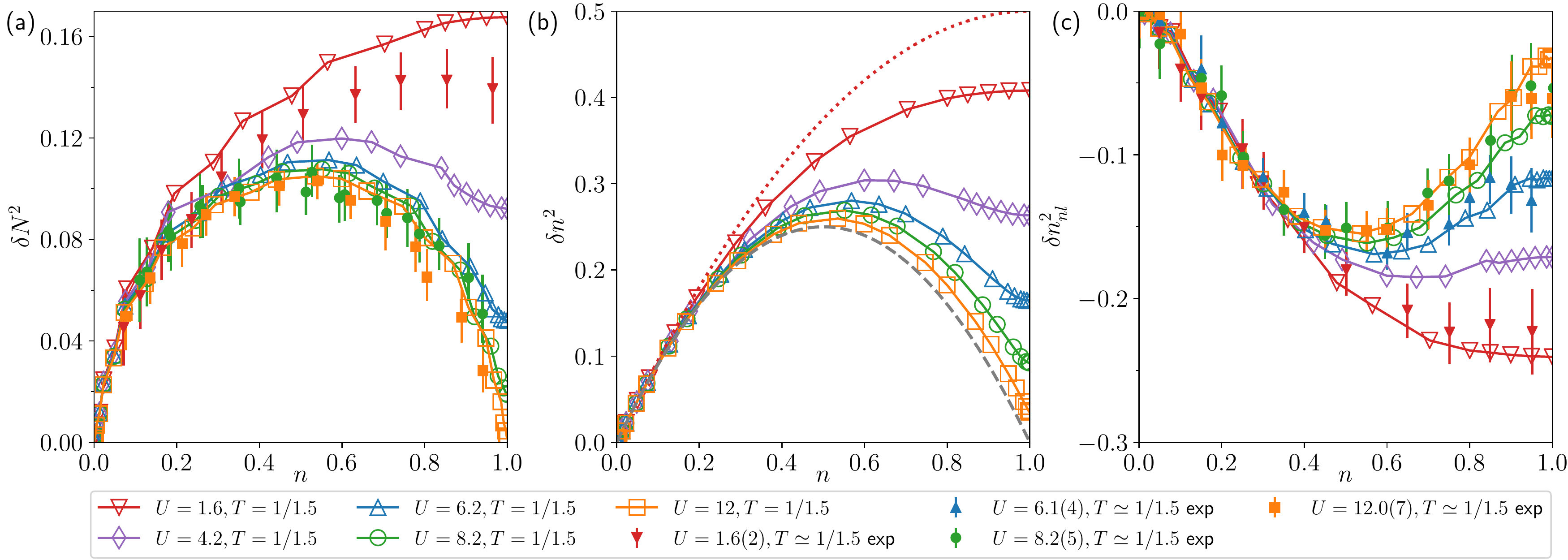}
}
\caption{(a) Thermodynamic density fluctuations $\delta N^2$ versus occupation $n$ at temperature $T=1/1.5$ for different values of the interaction strength $U$. Thermodynamic fluctuations can be decomposed into a local part, $\delta n^2$, plus a nonlocal part, $\delta n_{nl}^2$. (b) Local density fluctuations $\delta n^2$ versus $n$ for the same model parameters as in panel (a). Dotted red line and dashed gray line are exact results for $U=0$ and $U=\infty$, respectively. (c) Nonlocal density fluctuations $\delta n_{nl}^2$ versus $n$ for the same model parameters as in panel (a). Full symbols are experimental data with ultracold atoms in Ref.~\cite{DrewesPRL2016}. We work with units where $A=k_B=1$. 
}
\label{fig1}
\end{figure*}

\section{Model and method}
\label{S:Model}
We consider the one-band two-dimensional Hubbard model on a square lattice. Its Hamiltonian reads
\begin{equation}
H=-\sum_{\langle ij\rangle \sigma}t_{ij} c_{i\sigma}^\dagger c_{j\sigma}
  +U\sum_{i} n_{i\uparrow } n_{i\downarrow }
  -\mu\sum_{i\sigma} n_{i\sigma},  
\label{eq:HM}
\end{equation}
where $t_{ij}$ is the hopping between first neighbors (and $t_{ij}=1$ sets our energy units), $U$ is the repulsive on-site interaction, and $\mu$ is the chemical potential. The operators $c^{\dagger}_{i\sigma}, c_{i\sigma}$ respectively create and destroy a fermion with spin $\sigma=\{ \uparrow, \downarrow \}$ at site $i$, and $n_i=c_i^\dagger c_i$ is the number operator.
We solve this model using plaquette cellular DMFT~\cite{maier, kotliarRMP, tremblayR}. This method selects a cluster of lattice sites - here we consider a $2\times 2$ plaquette - and replaces the missing lattice environment with a self-consistent bath of noninteracting fermions. We solve the cluster in a bath problem with continuous-time quantum Monte Carlo in the hybridisation expansion (CT-HYB)~\cite{millisRMP}. 
Here we focus on the normal state only, where there is an equal number of fermions with spins up and down and the spin alignments have short range correlations but no long range order.   
We map out the phase diagram of Eq.~\ref{eq:HM} as a function of occupation $n$ and interaction strength $U$ for two decades in temperatures, from $T\approx 2$ down to $T \approx 1/100$. 
\footnote{Our CDMFT calculations required a large computation investment. At high temperature (e.g. $T=1/1.5$), one CDMFT iteration took about $20$ minutes on $48$ processors. At low temperature (e.g. $T=1/100$), the computational effort increases, and one CDMFT iteration took about 1 hour on $72$ processors. We normally reached convergence within $50$ CDMFT iterations, but close to phase boundaries and the Widom line hundreds of iterations were necessary. Once convergence is reached, averages were taken over at least the last $30$ CDMFT iterations (and up to hundreds close to phase boundaries). The resulting root mean square deviation on local quantities such as $n$ and $D$ that were needed to compute the density fluctuations were on the fifth digit.}

\section{Density fluctuations: definitions and comparison with available experiments}
\label{S:FluctHighT}

To extract the density fluctuations, we follow a protocol similar to the experimental work of Drewes et al.~\cite{DrewesPRL2016}. We first summarize this protocol by considering temperatures accessible to current experiments and by comparing our results to the experimental findings of Ref.~\cite{DrewesPRL2016}. Then in Sec.~\ref{S:FluctLowT} that follows, we extend our analysis to lower temperatures to discuss enhanced density fluctuations in proximity to a critical point at finite $T$ and $n$ which may be detected in future experiments.

\subsection{Thermodynamic fluctuations}

First, consider thermodynamic density fluctuations. The simplest measure of these fluctuations is the variance of the number of fermions $N$ in the area $A$: $\delta N^2 \equiv \langle (N - \langle N \rangle)^2 \rangle$. The fluctuation-dissipation theorem 
\begin{align}
\delta N^2 = A k_B  T n^2 \kappa_T 
\label{eq:FD-th}
\end{align}
links the thermodynamic fluctuations $\delta N^2$ to the charge compressibility $\kappa_T = \frac{1}{n^2} \left( \frac{\partial n}{\partial \mu} \right)_{T}$, where $n=N/A$ is the particle density. From the isothermal occupation curve $n(\mu)_T$ we extract $\kappa_T$ by numerical derivative, and hence using Eq.~\ref{eq:FD-th} we obtain $\delta N^2$. 

Figure~\ref{fig1}(a) shows the thermodynamic fluctuations $\delta N^2$ as a function of $n$ for different values of the interaction strength $U$. Our data are taken at temperature $T=1/1.5$, which is close to the lowest temperature that can be reached in current experiments. 

We point out three interesting features. 
First, there is excellent agreement between our calculations (open symbols) and the experimental results of Ref.~\cite{DrewesPRL2016} (full symbols). 
Second, the variance $\delta N^2$ is much smaller than expected from the density $n$ (i.e., $\delta N^2 / \langle N \rangle \ll 1$), implying sub-Poissonian statistics. This occurs even in the noninteracting case $U=0$. Indeed, as noticed in previous work~\cite{Muller:2010, Sanner:2010, DrewesPRL2016, Cheuk:2016}, this suppression of the thermodynamic density fluctuations at $U=0$ is a consequence of the Pauli exclusion principle. According to this principle, two fermions cannot occupy the same quantum state, which in real space means that fermions `avoid each other', or `antibunch', thereby suppressing the density fluctuations. This sharply differs from the ideal classical gas, for which $\kappa_T=1/(n k_B T)$, and hence, from the fluctuation-dissipation theorem Eq.~\ref{eq:FD-th}, $\delta N^2 / \langle N \rangle =1$, i.e. Poissonian statistics. 
Third, the suppression of thermodynamic density fluctuations $\delta N^2$ due to Pauli principle is further enhanced by interactions. Indeed, the interaction strength $U$ further reduces $\delta N^2$. As expected, the closer the occupation $n$ is to half filling $n = 1$, the larger the effect of $U$. 
Above the critical interaction $U_{\rm MIT} \sim 5.9$ where the system becomes an insulator at half-filling, the thermodynamic density fluctuations are strongly suppressed until they vanish completely at zero temperature, reflecting the incompressible nature of the Mott insulating state ($\kappa_T=0$).

\subsection{Local and nonlocal density fluctuations}

Next, we consider the length scales of the density fluctuations. From the definition of the occupation $n=n(\bf{x})$ in a translationally invariant system, one has~\cite{Zhou:2011}: 
\begin{equation}
k_B T \frac{\partial \langle n({\bf x}) \rangle}{\partial \mu} = \int d {\bf x'} \left[ \langle n({\bf x})n({\bf x'}) \rangle -\langle n({\bf x}) \rangle \langle n(\bf{x'}) \rangle \right]. 
\label{eq:DnDmu}
\end{equation}
Therefore, using the fluctuation-dissipation theorem Eq.~\ref{eq:FD-th}, the density fluctuations at all length scales are included in $\delta N^2$. Eq.~\ref{eq:DnDmu} thus suggests to write the thermodynamic fluctuations $\delta N^2$ as the sum of local fluctuations $\delta n^2$ and nonlocal fluctuations $\delta n_{nl}^2$~\cite{Duchon:2012, DrewesPRL2016}: $\delta N^2 = \delta n^2 + \delta n_{nl}^2$. For our two-dimensional lattice, Eq.~\ref{eq:DnDmu} reads: 
\begin{align}
n^2 \kappa_T &= \frac{1}{A k_B T} \sum_{ij} \left( \langle n_{i} n_{j}\rangle - \langle n_{i} \rangle \langle n_{j} \rangle \right) \nonumber \\
& = \frac{1}{A k_B T} \left[ \delta n^2 +\sum_{i\neq j} \left( \langle n_{i} n_{j}\rangle - \langle n_{i} \rangle \langle n_{j} \rangle \right)  \right], 
\end{align}
so 
\begin{align}
\delta N^2 = A k_B T n^2 \kappa_T = \delta n^2 + \delta n_{nl}^2,
\end{align}
where we have identified $\delta n_{nl}^2$ with $\sum_{i\neq j} \left( \langle n_{i} n_{j}\rangle - \langle n_{i} \rangle \langle n_{j} \rangle \right)$. Now the local (onsite) density fluctuations $\delta n^2$ can be calculated explicitly:
\begin{align}
\delta n^2  \equiv \langle n_i^2 \rangle - \langle n_i \rangle^2 = n + 2D - n^2, 
\label{eq:LocalFlu}
\end{align}
where $n=n_i=\langle n_{i\uparrow} \rangle + \langle n_{i\downarrow} \rangle$ is the density, $\langle n_{i\uparrow} \rangle = \langle n_{i\downarrow} \rangle$ is the spin-resolved density, and $D=\langle n_{i\uparrow}n_{i\downarrow} \rangle$ is the double occupancy. So, the nonlocal fluctuations
\begin{align}
\delta n_{nl}^2 = \delta N^2 - \delta n^2 = A k_B T n^2 \kappa_T - \delta n^2
\label{eq:NlFlu}
\end{align}
can be obtained as the difference between the thermodynamic and local fluctuations.

Figure~\ref{fig1}(b) and Figure~\ref{fig1}(c) show the local and nonlocal fluctuations $\delta n^2$ and $\delta n_{nl}^2$ respectively as a function of $n$ for the same model parameters as Fig.~\ref{fig1}(a). Again our calculations (open symbols) and the experimental data of Ref.~\cite{DrewesPRL2016} (full symbols) are in excellent agreement, as shown in Figure~\ref{fig1}(c). 
The behavior of the local (onsite) fluctuations $\delta n^2$ in Figure~\ref{fig1}(b) can be understood by considering the $U=0$ and the $U=\infty$ limits. For $U=0$, $D=n^2/4$, so $\delta n^2 = n-n^2/2$ (red dotted line), which is a monotonically increasing function of $n$. For $U=\infty$, $D=0$, so $\delta n^2 = n-n^2$ (gray dashed line), which is a non-monotonic function of $n$. Therefore, the effect of $U$ is to strongly reduce the local density fluctuations, especially close to half filling.

By comparing the overall magnitude of $\delta N^2$ in Figure~\ref{fig1}(a) and $\delta n^2$ in Figure~\ref{fig1}(b), we find that the inequality $\delta N^2 \ll \delta n^2$ is satisfied over the whole range of fillings $n$: clearly, the local fluctuations $\delta n^2$ do not account for all thermodynamic fluctuations, which also comprise nonlocal fluctuations $\delta n_{nl}^2$ that are negative and of comparable magnitude. Fig.~\ref{fig1}(c) shows the filling dependence of $\delta n_{nl}^2$ calculated with Eq.~\ref{eq:NlFlu}. These negative contributions to $\delta N^2$ lead to sub-Poissonian statistics. As described in previous work~\cite{Cheuk:2016, Cheuk:thesis}, the behavior of $\delta n_{nl}^2$ is controlled by the competing effects of Fermi statistics, which dominates for low density and small values of $U$, and superexchange, which conversely dominates for large values of $U$ and upon approaching half filling. 

More specifically, for $U=0$, nonlocal correlations arise from the Pauli principle that keeps apart fermions with the same spin, so fermions move in a way to `avoid each other', or antibunch. This causes a suppression in charge density in the vicinity of any fermion, and this charge suppression can be thought of as a hole, and is usually called exchange or Pauli hole. At low density, the interaction strength $U$ can enhance this antibunching effect, giving rise to what is known as a correlation hole~\cite{MahanBook}. 
As a result, from low to moderate density, $n < 0.5$, $\delta n_{nl}^2$ grows in absolute value, i.e. becomes more negative, with increasing density since there are more fermions in the lattice and thus the average fermion spacing decreases with increasing $n$ and the antibunching effect becomes more pronounced. In that range of filling, $\delta n_{nl}^2$ does not vary much with increasing $U$. 

By contrast, from moderate density up to half filling, $n>0.5$, $U$ has an appreciable effect. We see in Figure~\ref{fig1}(c) that as $U$ increases, $\delta n_{nl}^2$ decreases in absolute value, i.e. become less negative. This is because upon approaching half filling $n=1$, the interaction strength $U$ reduces the double occupancies, so the fermions become more localised around their sites, meaning that the charge becomes frozen. In that region near half-filling, and at large $U$, another way to understand negative non-local correlations is to realize that they contribute to lower the energy through superexchange $J$. Indeed, $J$ comes from virtual double occupation produced by a fermion leaving a hole behind and hopping on a site occupied by a fermion of opposite spin. Effectively, this creates correlations between empty sites (holons) and doubly occupied sites (doublons) on neighboring sites~\footnote{Virtual processes occur across a bond between neighboring sites of the lattice. It is intriguing that the virtual process description becomes consistent near  $n=0.5$, the classical bond-percolation threshold of the two-dimensional square lattice.}. Holons and doublons are then close together, or `bunched', i.e. they form quasi-bound states that lead to negative near-neighbor correlations as can be seen from the definition of nonlocal fluctuations: when there is a virtual double occupancy, then a neighboring site will be empty and the product $n_{i} n_{j}$ will be zero so that the contribution of that process to $\sum_{i\neq j} \left( \langle n_{i} n_{j}\rangle - \langle n_{i} \rangle \langle n_{j} \rangle \right)$ will be negative. However, since $J=4t^2/U$ is inversely proportional to $U$, the larger the value of $U$, the weaker this doublon-holon correlation since there is no more energy gain associated with virtual hops. Eventually then, non-local fluctuations decrease with increasing $U$. 
In the Mott insulating state at large $U$ and $n=1$, thermodynamic fluctuations essentially become local fluctuations. 
This description is complementary to the explanation in Refs.~\cite{Cheuk:2016, Cheuk:thesis} and Ref.~\cite{DrewesPRL2016}. The doublon-holon correlations are also important in spin transport~\cite{Nichols:2018}.

\subsection{Temperature dependence of nonlocal density fluctuations}

\begin{figure}%[h!]
\centering{
\includegraphics[width=1.\linewidth]{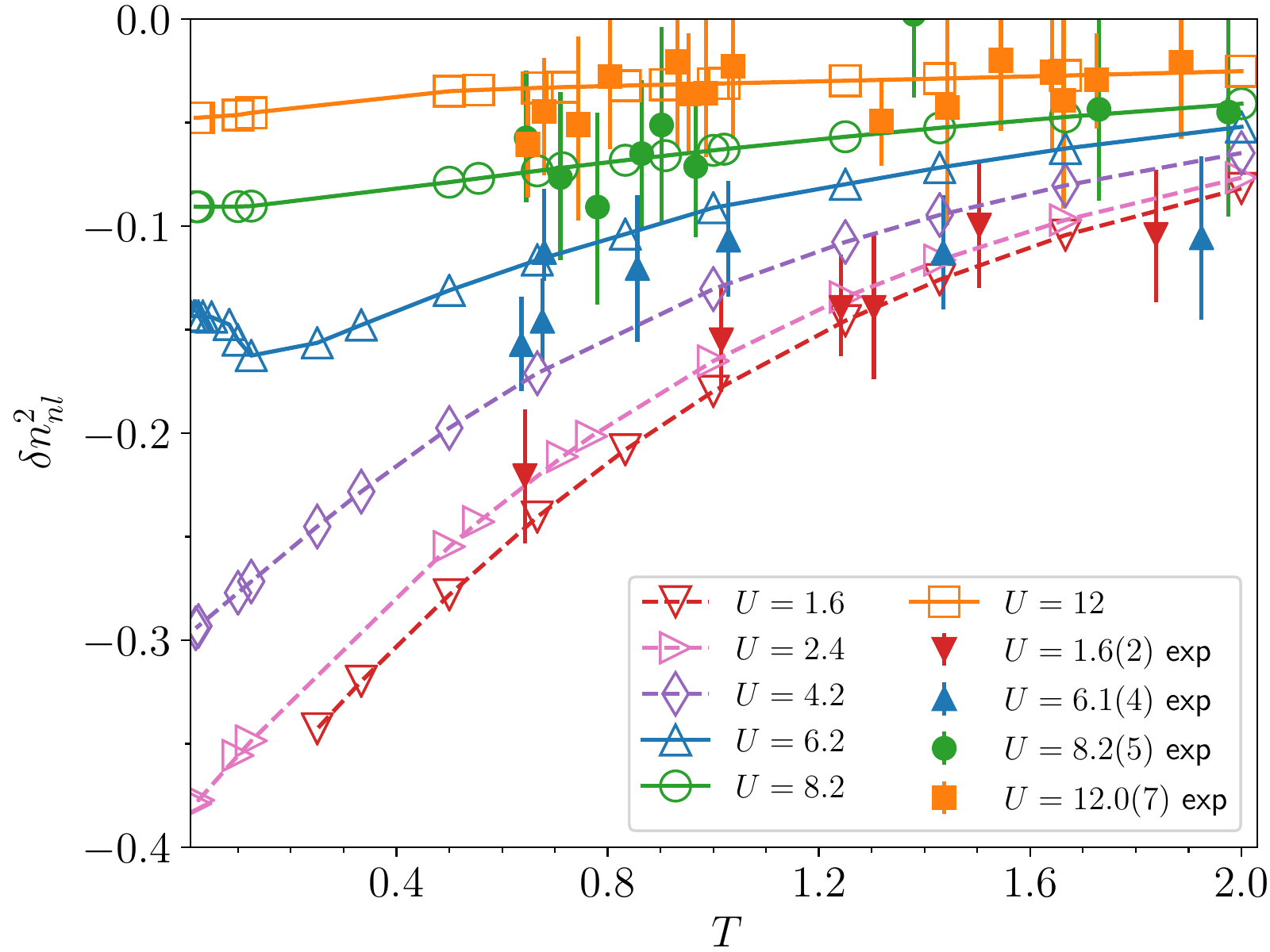}
}
\caption{Nonlocal density fluctuations $\delta n_{nl}^2$ versus temperature $T$ for different values of $U$, at half filling $n=1$. Full symbols are experimental data with ultracold atoms of Ref.~\cite{DrewesPRL2016}.
}
\label{fig2}
\end{figure}

Up to now we considered one temperature only, $T=1/1.5$, which is close to the lowest temperature reached by ultracold atom experiments in Ref.~\cite{DrewesPRL2016}. Further insight can be gained from the temperature evolution of the nonlocal fluctuations $\delta n_{nl}^2$. Figure~\ref{fig2} shows $\delta n_{nl}^2$ as a function of temperature for different values of $U$, at half-filling $n=1$. An important finding of Ref.~\cite{DrewesPRL2016} is that nonlocal fluctuations decrease in absolute value upon increasing $T$ (see experimental data in our Fig.~\ref{fig2}). Physically, this is because with increasing $T$ the De Broglie thermal wavelength $\xi_{th}$ becomes smaller than the average distance between fermions, freezing nonlocal quantum fluctuations, hence recovering classical Poissonian statistics. 

Our results in the temperature range $1/1.5<T<2$, displayed in Fig.~\ref{fig2}, confirm this behavior and are in good quantitative agreement with experimental data (solid symbols). Furthermore, in our CDMFT simulations we can access lower temperatures than those experimentally available. In a metallic state, for $U<U_{\rm MIT}$, $\delta n_{nl}^2$ remains a monotonic function of $T$ at all temperatures and thus becomes more negative with decreasing $T$. In sharp contrast, for $U>U_{\rm MIT}$, a new temperature scale emerges where $\delta n_{nl}^2$ becomes less negative or saturates at low temperature. This occurs because the system is entering the Mott insulating state as the temperature is reduced. Hence, states become more localized, or from another point of view, the doublon-holon correlations lead to a reduction of the magnitude of the nonlocal correlations. 

Far away from half filling, $n \le 0.75$, the system is metallic and Ref.~\cite{DrewesPRL2016} demonstrates that $\delta n_{nl}^2$ is a monotonic function of $T$, as for the case $U<U_{\rm MIT}$ at $n=1$ shown in Fig.~\ref{fig2}. Thus in a metallic state it is the De Broglie thermal wavelength $\xi_{th}$ that controls the temperature behavior of  $\delta n_{nl}^2$. 
In the next section we show that, for $U>U_{\rm MIT}$ and $n>0.75$ (so in a doped Mott insulator), a novel emergent phenomenon due to the interplay of Mott physics and short-range correlations is controlling the low temperature behavior of $\delta n_{nl}^2$.

\section{Density fluctuations across the doping driven Mott transition} 
\label{S:FluctLowT}

In this section we analyze the behavior of the density fluctuations at temperatures lower than those reached by current experiments. 

\begin{figure*}%[h!]
\centering{
\includegraphics[width=1.\linewidth]{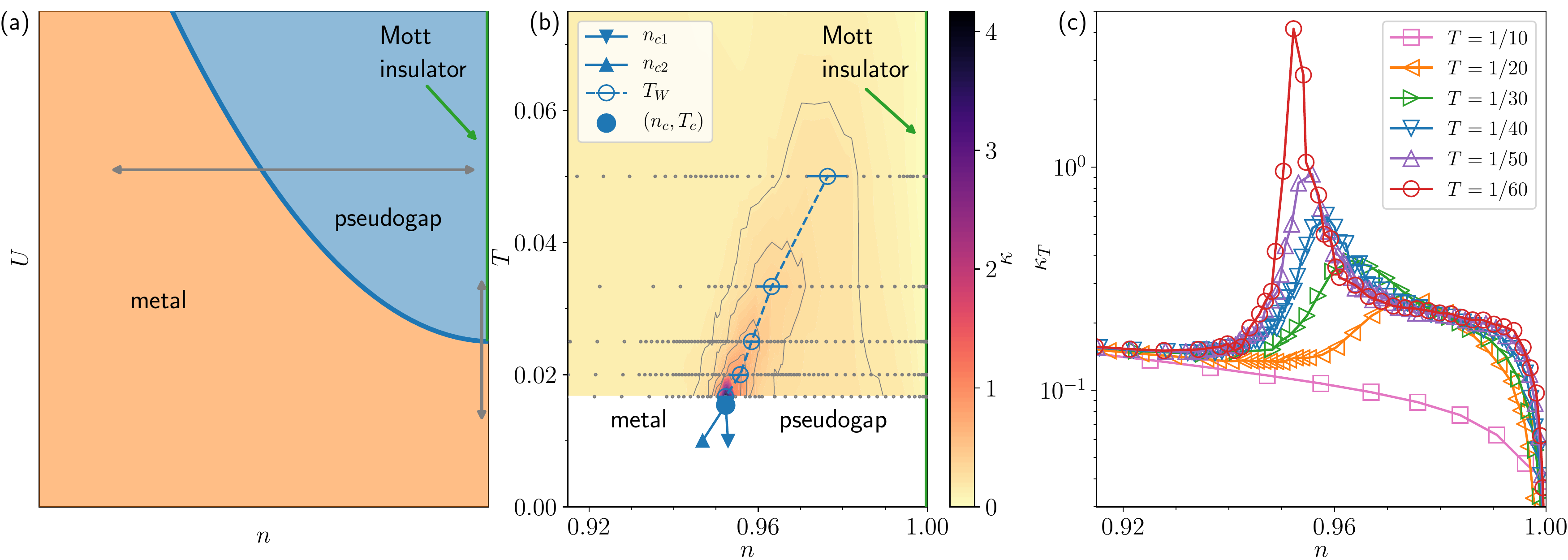}
}
\caption{(a) Sketch of the low-temperature $U-n$ phase diagram of the two-dimensional Hubbard model in the normal state solved with plaquette CDMFT~\cite{sht, sht2}. The defining feature of the model is the blue line, marking the first-order transition between a metal (orange region) and Mott insulator at $n=1$ (green vertical line at $n=1$), and between a metal and a pseudogap (blue region) for $n\neq 1$. The former refers to the $U$-driven Mott transition (see vertical double arrow), the latter denotes the doping (or filling $n$) driven Mott transition (see horizontal double arrow). (b) Color map of the charge compressibility $\kappa_T$ in the $T-n$ phase diagram for $U=6.2>U_{\rm MIT}$. The first-order transition is bounded by the spinodal lines $n_{c1}$ and $n_{c2}$ (lines with solid triangles). The region between $n_{c1}$ and $n_{c2}$ is an instability region. The first-order transition ends in a critical endpoint at $(n_c, T_c)$, from which a line of a rapid crossover emerges (the Widom line, $T_W$, marked by open circles). (c) Semi-logarithmic plot of isothermal charge compressibility $\kappa_T$ versus $n$ at $U=6.2$ for several temperatures. The maxima of the compressibility define the Widom line appearing in panel (b).
}
\label{fig3}
\end{figure*}
\begin{figure*}%[h!]
\centering{
\includegraphics[width=1.\linewidth]{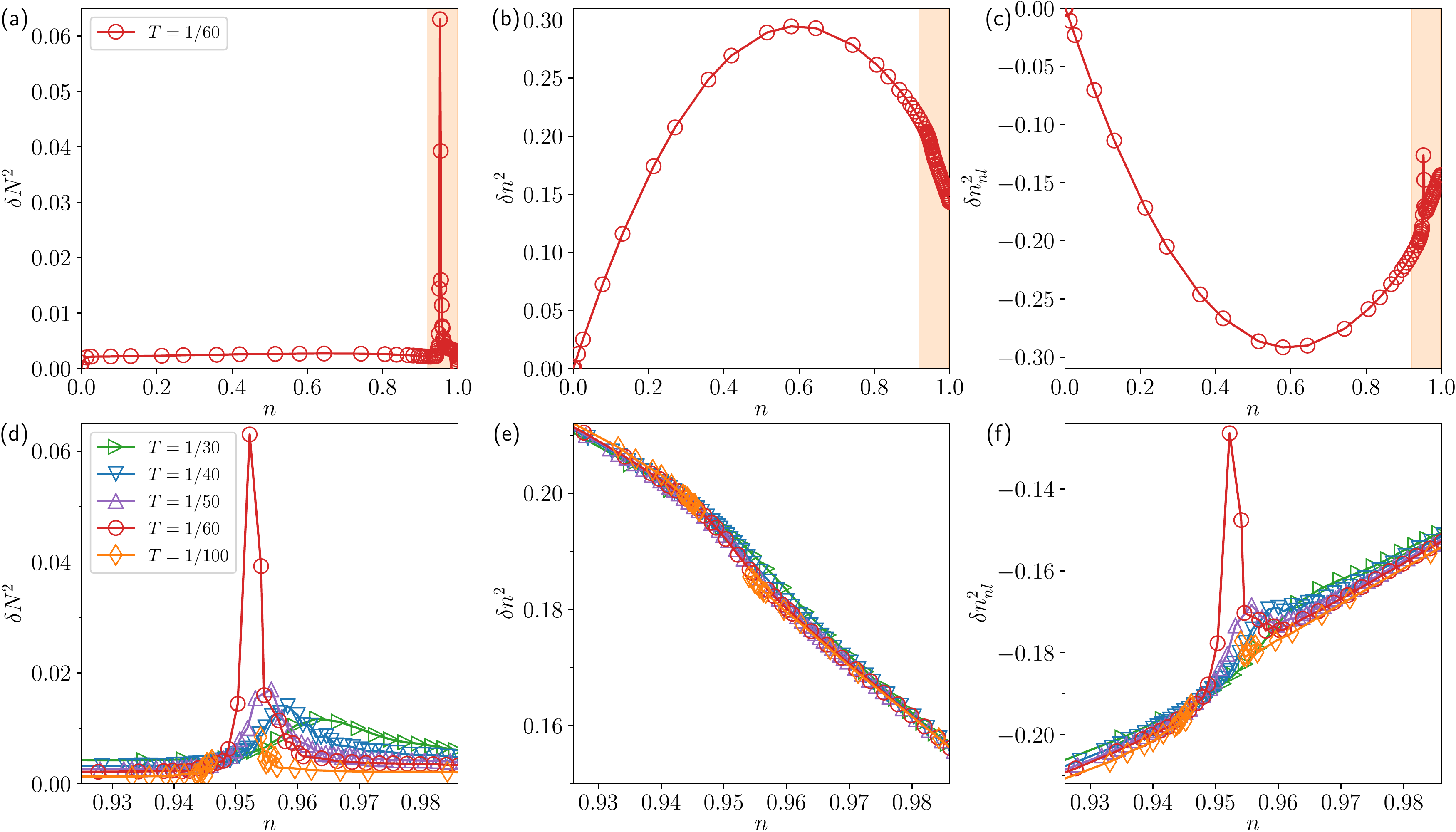}
}
\caption{(a) Thermodynamic density fluctuations $\delta N^2$ versus $n$ for $U=6.2 > U_{\rm MIT}$ at temperature $T=1/60$, which is close to the critical endpoint temperature $T_c$. (b) Local density fluctuations $\delta n^2$ versus $n$ for $U=6.2$ and $T=1/60$. (c) Nonlocal fluctuations $\delta n_{nl}^2$ versus $n$ for $U=6.2$ and $T=1/60$. Panels (d), (e), (f) are a zoom into the shaded orange regions of panels (a),(b),(c), for several temperatures, $T=1/30, 1/40, 1/50, 1/60$ which lie above $T_c$ and for $T=1/100$ which lies below $T_c$ (and thus it is discontinuous as a function of $n$).  
}
\label{fig4}
\end{figure*}

\subsection{First-order transition, critical point, and Widom line}

According to previous theoretical work with plaquette CDMFT~\cite{sht, sht2}, the key qualitative feature of the $U-n$ phase diagram of the normal state two-dimensional Hubbard model is a low-temperature first-order transition extending from half filling $n=1$ to away from half filling, as sketched with a blue line in the $U-n$ phase diagram in Figure~\ref{fig3}(a). 
At $n=1$ this transition separates a metal from a Mott insulator as $U$ increases (this corresponds to the familiar $U$ driven Mott transition). For $n\neq 1$ this transition separates two metallic states: as $n$ is lowered, one moves from a strongly correlated metal with a pseudogap to a correlated metal. 

Adding the temperature axis, the first-order transition line in the $U-n$ phase diagram of Fig.~\ref{fig3}(a) becomes a surface of first-order transition in the $U-n-T$ space. Upon increasing $T$, this surface ends in a second order critical line, from which a crossover emerges. A slice of the $U-T-n$ space at constant $U$, with $U>U_{\rm MIT}$, thus gives rise to the $T-n$ phase diagram of Fig.~\ref{fig3}(b), which contains our CDMFT results for $U=6.2$ (i.e. a value of the interaction $U$ slightly larger than the critical threshold for obtaining a Mott insulator at $n=1$). The first-order transition is delimited by the spinodal lines $n_{c1}$ and $n_{c2}$ (line with solid triangles), where the pseudogap phase and the metallic phase cease to exist, respectively. This first-order transition ends in a second order critical endpoint at $(n_c, T_c)\sim (0.95, 0.015)$ ($T_c\sim 0.004$ in units of the bare half-bandwidth), from which a rapid crossover, known as Widom line~\cite{water1, supercritical}, emerges. The endpoint is defined as the location of the divergence of the correlation length and the Widom line as the locus of maxima of the correlation length. As $U$ increases, the critical temperature $T_c$ decreases rapidly while the corresponding critical filling $n_c$ increases. While the first-order transition may be masked by long-range ordered phases at low temperature, one expects that in frustrated systems the Widom line will continue to describe high-temperature crossovers. 

Operationally we construct the Widom line with the local maxima of the isothermal charge compressibility $\kappa_T$ as a function of $n$ (blue dashed line in Fig.~\ref{fig3}(b)). As shown in Fig.~\ref{fig3}(c), the rather featureless $\kappa_T(n)$ at high temperature develops a broad peak that narrows and grows rapidly in intensity as $T$ decreases towards $T_c$, where $\kappa_T$ diverges. Hence, a region of enhanced charge compressibility emanates from the endpoint into the supercritical region $T>T_c$ (see Fig.~\ref{fig3}(b) and Fig.~\ref{fig3}(c)), all the way up to $T \approx 5T_c$~\cite{ssht}. Crossing the Widom line along isotherms or along paths at constant $n$ produces sharp but continuous changes in the thermodynamic~\cite{ssht} properties, such as compressibility, dynamic~\cite{ssht} properties such as the local density of states, and transport properties such as $c$-axis resistivity~\cite{sshtRHO}. In the following, we focus on the changes of the density fluctuations upon crossing the Widom line.

\subsection{Critical opalescence}

A defining feature of the critical point is the divergence of the correlation length $\xi$, which quantifies the characteristic distance over which fermion-density fluctuations are correlated. The divergence of $\xi$ is linked to the development of fluctuations in the density at all scale up to infinity, leading to the well known phenomenon of critical opalescence: physically, in our case, this means that any two fermion-density fluctuations are correlated, no matter the distance between them. Thus the thermodynamic fluctuations $\delta N^2$ are expected to diverge. The fluctuation-dissipation theorem associates the divergence of $\delta N^2$ with the divergence of $\kappa_T$. 
This expectation is fully confirmed by the calculation of $\delta N^2(n)$ for different temperatures at $U=6.2$, as shown in Figs.~\ref{fig4}(a)(d). Away from half-filling, in the region $0.92<n<1$ [shaded region in Fig.~\ref{fig4}(a)], the thermodynamic fluctuations $\delta N^2$ show an anomalous peak that narrows in size and increases in magnitude as the endpoint $T_c$ is approached from above (see Fig.~\ref{fig4}(d)). 
It extends up to $T \approx 1/14$ or $T \approx 5 T_c$. The position of these peaks are the same of those defining the Widom line. Detection of this non monotonic behavior is a precursor of the endpoint. This behavior is not surprising since we used the fluctuation-dissipation theorem Eq.~\ref{eq:FD-th} to calculate $\delta N^2$, and $\kappa_T$ diverges at $T_c$ (Fig.~\ref{fig3}(b) and Ref.~\cite{ssht} for a detailed discussion).

\subsection{Local and nonlocal fluctuations near criticality}

Since the pioneering work of Kadanoff~\cite{Kadanoff:1966} and Wilson~\cite{Wilson:1971a, Wilson:RMP1975}, it is well known that at the critical point there are fluctuations at {\it all} length scales: at $T_c$ the scale of the largest density fluctuations becomes infinite, however fluctuations at smaller scales persist, as illustrated by the phenomenon of critical opalescence in fluids. What remains interesting then, is to pin down the contribution of different length scales of the thermodynamic density fluctuations. Here we focus on the relative contributions of local and of nonlocal fluctuations to the thermodynamic fluctuations $\delta N^2$. 

Let us begin with the local fluctuations $\delta n^2$. Figure~\ref{fig4}(b) shows $\delta n^2$ versus $n$ for $U=6.2$ near $T_c$. In sharp contrast with $\delta N^2$, $\delta n^2(n)$ does not show an enhancement along the Widom line (see Fig.~\ref{fig4}(e)). Its temperature dependence is controlled by that of double occupancy so it is very weak when $T$ is much less than $U$. 
Local fluctuations $\delta n^2$ are mostly dominated by quantum effects, as we discuss momentarily.  

Next we turn to the nonlocal fluctuation $\delta n_{nl}^2$, shown in Figure~\ref{fig4}(c) again for a temperature close to $T_c$. Upon crossing the Widom line, $\delta n_{nl}^2$ has a non-monotonic behavior, showing an additional upward peak as a function of $n$, which sharpens and narrows as the critical endpoint is approached (see Fig.~\ref{fig4}(f)). Hence, quite surprisingly, $\delta n_{nl}^2$ becomes {\it less negative} along the Widom line. Note that away from the Widom line (say for $n<0.9$), $\delta n_{nl}^2$ becomes more negative with decreasing $T$, in agreement with the discussion of Figure~\ref{fig1}(c).  

To understand the behavior of $\delta n_{nl}^2$ and of $\delta n^2$ one needs to realize that close to the endpoint and to the Widom line, the system enters into the so-called renormalized classical regime~\cite{RCR1989}. In this regime, the fluctuations are large and the correlation length $\xi$ becomes larger than the thermal De Broglie wavelength $\xi_{th}$, and hence the most important fluctuations are classical. A large correlation length is associated with a small energy scale so, equivalently, in the renormalized classical regime the characteristic energy of the fluctuations becomes smaller than the energy associated to temperature ($\hbar \omega < k_B T$). 
Close to $T_c$ and to the Widom line, the correlation length $\xi$ grows rapidly (equivalently, there is ``critical slowing down'' so the characteristic frequency of the system decreases), and hence the thermodynamic fluctuations $\delta N^2$ are dominated by the classical contributions ($\chi(q=0,0) \sim \xi^2$ in mean-field). By contrast with the quantum contribution to the nonlocal fluctuations $\delta n_{nl}^2$ that is negative, the classical (thermal) contribution to the nonlocal fluctuations $\delta n_{nl}^2$ is {\it positive}. Hence $\delta n_{nl}^2$ becomes less negative close to $T_c$ or to the Widom line. Returning to the local fluctuations $\delta n^2$, the classical (i.e. thermal) contribution to $\delta n^2$ is logarithmic in $\xi$, as one can see using the Ornstein-Zernike form for the susceptibility $(\int_0^\Omega \chi(q,0) d^2q \sim \int_0^\Omega \frac{A}{\xi^{-2} +q^2} d^2q \sim A \ln(\Omega/\xi))$. This means that $\delta n^2$ does not depend much on $\xi$. It is controlled mainly by quantum effects, even though strictly speaking there is a diverging contribution at $T_c$.  

Therefore there are two mechanisms governing the temperature behavior of $\delta n_{nl}^2$ {\it away from half-filling}, $n<1$, for a doped Mott insulator.  
The first mechanism occurs at high temperature, as discussed in Sec.~\ref{S:FluctHighT}. As the temperature decreases below the Fermi energy, quantum mechanical effects become important when $\xi_{th}$ becomes larger than the average distance between fermions $\ell$ and it is $\xi_{th}$ then that sets a limit to the range of density fluctuations. The non-local correlations are negative because of the Pauli principle and because of fermion-fermion repulsion.  
A second mechanism occurs at much lower temperatures: as $T$ crosses the Widom line, or the critical endpoint, the system becomes renormalized classical: $\xi$ becomes larger than $\xi_{th}$ and it is $\xi$ that sets a limit to the range of density fluctuations. Since classical contribution to $\delta n_{nl}^2$ are positive, $\delta n_{nl}^2$ becomes less negative close to $T_c$ or to the Widom line.

\section{Predictions for experiments}
\label{S:Prediction}

Our results lead to the following predictions for quantum simulations of the two-dimensional Hubbard model with ultracold atoms in optical lattices. For $U$ larger than the critical value $U_{\rm MIT}$ where the Mott transition occurs at half-filling, both the charge compressibility $\kappa_T$ and the thermodynamic density fluctuations $\delta N^2$ will show a peak as a function of the occupation $n$. As $T$ is decreased toward $T_c$, the peak will increase in magnitude and narrow in size. This enhancement in $\delta N^2$ is associated with the nonlocal contribution $\delta n_{nl}^2$ becoming less negative, while the local fluctuations $\delta n^2$ do not do much. These will be the signature of the Widom line. Below $T_c$, a first-order transition occurs at finite density and finite temperature, and hence $\delta N^2$ will be discontinuous. 

Usual provisos apply. First, the presence of long-range ordered phases, or quasi-long-range ordered phases such as antiferromagnetism~\cite{mazurenko2017cold, LorenzoAF} or superconductivity~\cite{sshtSC, LorenzoSC} in the low temperature region of the $T-n$ phase diagram can prevent the observation of the endpoint or of some portion of the Widom line. Nevertheless, the Widom line extends far away from the endpoint, up to $T \approx 5T_c$, so it may come out into the normal state. Furthermore, it is possible to shrink the regime of competing states, such antiferromagnetism, by adding frustration (e.g next nearest neighbor hopping $t'$~\cite{LorenzoAF2}, or triangular lattice~\cite{Hebert:2015}). The onset temperature $T_N$ of the antiferromagnetic correlations strongly depends on the interaction $U$ and occupation $n$. Within plaquette CDMFT, it is largest at $n=1$ (i.e. zero doping) around $U \approx 7$, where $T_N \approx 0.33$~\cite{LorenzoAF}. It decays rapidly as a function of $U$ and $n$. For instance, for $U=6.2$ considered in Sec.~\ref{S:FluctLowT}, antiferromagnetism sets in from $n=1$ to $n \approx 0.85$~\cite{LorenzoAF2}, similarly to recent experimental data with ultracold atoms in the antiferromagnetic phase~\cite{mazurenko2017cold}. Therefore the results of this work at half filling $n=1$ of Fig.~\ref{fig2} are quantitatively correct down to at least $T\approx 0.33$. Away from half filling, the results for $U=6.2$ in Figs.~\ref{fig3},\ref{fig4} are quantitatively correct for $n \lessapprox 0.85$, but only qualitative for $n \gtrapprox 0.85$.

Second, our results have been obtained with plaquette CDMFT that treats exactly local and short-range fluctuations within the plaquette. The fate of the first order transition and its associated Widom line with larger cluster size including larger range of fluctuations is computationally very challenging and hence has not been explored systematically. Nevertheless, one of our key findings is that nonlocal fluctuations {\it decrease} in absolute value near the Widom line and close to $T_c$. This suggests that larger nonlocal correlations probed by larger clusters should not qualitatively modify our results. 

The significance of the Widom line emanating from the endpoint at finite $T$ and $n$ stems from the theoretical proposal~\cite{ssht} that identifies the Widom line with the rapid crossovers marking the pseudogap temperature $T^*$ in the phase diagram of hole-doped cuprates~\cite{ssht, sshtRHO}. If confirmed by experiments with ultracold atoms, this result could throw new light into the explanation of the pseudogap phase in cuprate high-temperature superconductors.

\begin{acknowledgments}
This work has been supported by the Natural Sciences and Engineering Research Council of Canada (NSERC) under grants RGPIN-2014-04584, the Canada First Research Excellence Fund and by the Research Chair in the Theory of Quantum Materials. Simulations were performed on computers provided by the Canadian Foundation for Innovation, the Minist\`ere de l'\'Education des Loisirs et du Sport (Qu\'ebec), Calcul Qu\'ebec, and Compute Canada.
\end{acknowledgments}

%\bibliography{f}

%merlin.mbs apsrev4-1.bst 2010-07-25 4.21a (PWD, AO, DPC) hacked
%Control: key (0)
%Control: author (0) dotless jnrlst
%Control: editor formatted (1) identically to author
%Control: production of article title (0) allowed
%Control: page (1) range
%Control: year (0) verbatim
%Control: production of eprint (0) enabled
%

\end{document}